\documentclass[aps,twocolumn,showpacs,showkeys,superscriptaddress,amsmath,amssymb,nofootinbib,floatfix,prc]{revtex4-1}

\usepackage{graphicx}
\usepackage{subfigure}
\usepackage{bm}

\makeatletter

\begin{document}

\title{Collective dynamics of the high-energy proton-nucleus collisions}

\author{Piotr Bo\.zek}
\email{Piotr.Bozek@ifj.edu.pl}
\affiliation{AGH University of Science and Technology, Faculty of Physics and Applied Computer Science, al. Mickiewicza 30, 30-059 Krakow, Poland}
\affiliation{The H. Niewodnicza\'nski Institute of Nuclear Physics, Polish Academy of Sciences, PL-31342 Krak\'ow, Poland}

\author{Wojciech Broniowski}
\email{Wojciech.Broniowski@ifj.edu.pl} 
\affiliation{The H. Niewodnicza\'nski Institute of Nuclear Physics, Polish Academy of Sciences, PL-31342 Krak\'ow, Poland} 
\affiliation{Institute of Physics, Jan Kochanowski University, PL-25406~Kielce, Poland}
\affiliation{CNRS, URA2306, Institut de Physique Th\'eorique de Saclay, F-91191 Gif-sur-Yvette, France}


\begin{abstract}

We analyze the proton-lead collisions at the LHC energy of $\sqrt{s_{NN}}=5.02$~TeV in the three-stage approach, previously 
used to successfully describe the relativistic A-A collisions. The approach consists of the early phase, modeled with 
the Glauber model, the event-by-event viscous 3+1~dimensional (3+1~D) relativistic hydrodynamics, and the statistical hadronization at freeze-out. 
We show that features typical of collective dynamics, such as the harmonic flow and the ridge structures in the two-particle 
correlations in relative azimuth and pseudorapidity, may be naturally explained in our framework. In the proton-nucleus system 
the harmonic flow is generated from an initially event-by-event deformed system and is entirely due to these initial fluctuations. 
Notably, fluctuations of strength of the initial Glauber sources which yield the observed distribution of hadron multiplicities and,
at the same time, lead to correct values of the elliptic flow coefficients both from the two- and four-particle cumulant method, as measured
by the ATLAS collaboration. The azimuthally asymmetric flow is not modified significantly when changing 
the  viscosity coefficient, the initial time for the collective expansion, or the initial size of the fireball. 
The results present an estimate of the collective component in the two-particle correlations measured experimentally. 
We demonstrate that the harmonic flow coefficients can be experimentally measured with methods based on large rapidity gaps which reduce 
some of the other sources of correlations.
\end{abstract}



\maketitle

\section{Introduction \label{sec:intro}}

The recent interest in proton-nucleus (p-A) collisions stems from the expectations that 
the experimental data for this system could be used to test 
various theoretical approaches developed for relativistic collisions~\cite{Salgado:2011pf}, moreover, 
it could serve as a reference for experiments involving 
nucleus-nucleus (A-A) collisions. An interesting possibility is that the 
collective behavior clearly seen in the  A-A collisions  may be present 
already in the p-A collisions, and even in the proton-proton (p-p) collisions of highest 
multiplicity of the produced particles. 
The experimental~\cite{ALICE:2012xs,ALICE:2012mj,CMS:2012qk,Abelev:2012cya,ATLAS:pPb,Adare:2013piz,Khachatryan:2010gv} 
and  theoretical~\cite{Luzum:2009sb,*d'Enterria:2010hd,*Bozek:2009dt,*CasalderreySolana:2009uk,*Avsar:2010rf,Bozek:2010pb,*Werner:2010ss,Bozek:2011if,Bozek:2012gr,Shuryak:2013ke} 
investigations can provide a limit on the amount of the collective flow in small systems,  
setting a boundary on the collective behavior and helping to answer the questions: 
How small may the system be and under what 
conditions it is still describable with hydrodynamics? What observables are most sensitive to 
the collectivity? What is the interplay of various stages of the dynamics, starting from the 
initial condition, through the intermediate evolution, to hadronization? 

The studies of the A-A collisions at RHIC~\cite{Arsene:2004fa,*Adcox:2004mh,*Back:2004je,*Adams:2005dq}
and the LHC~\cite{Collaboration:2011yba,*Li:2011mp,*collaboration:2011hfa} led to by now 
conventional interpretation of numerous observed phenomena in the mid-rapidity 
region via formation of a hot dense medium -- the strongly interacting quark-gluon plasma -- which 
evolves as a fluid and may be successfully described with relativistic 
hydrodynamics~\cite{Teaney:2009qa,*Broniowski:2008vp,*Shen:2011zc,*Schenke:2011zz,*Luzum:2011mm,*Ollitrault:2012cm,*newreview}.
Basic phenomena supporting this view are
\begin{itemize}
 \item The harmonic flow (elliptic, triangular, higher-order) 
\cite{Ollitrault:1992,Alver:2010gr,Arsene:2004fa,Collaboration:2011yba}
 \item Characteristic ridge structures seen ~\cite{Alver:2008aa,*Abelev:2009af,*ATLAS:2012at,*Aamodt:2011by,*Agakishiev:2011pe,*Chatrchyan:2011eka,CMS:2012qk,Li:2011mp}
 in the 2-particle correlation functions in relative pseudorapidity and azimuth, possible to explain with 
 harmonic flow~\cite{Voloshin:2004th,*Takahashi:2009na,*Luzum:2010sp,*Sorensen:2011hm,*Werner:2012xh,Bozek:2012en}.
 \item Specific features of the interferometric radii~\cite{Abelev:2009tp,Aamodt:2011mr}.
 \item Jet quenching (see, e.g.,~\cite{Betz:2012hv} and references therein).
\end{itemize}
In the study presented in this article we investigate the first two items from the 
above list for the case of the p-Pb collisions, recently studied at the LHC at the collision
energy of $\sqrt{s_{NN}}=5.02$~TeV. We  apply a 
treatment based on hydrodynamics to find quantitative estimates for the measured quantities, 
extending an early analysis of flow~\cite{Bozek:2011if} by one of us (PB) in the p-A and deuteron-nucleus 
systems, as well as the more recent event-by-event 
studies of the correlation functions~\cite{Bozek:2012gr} and the interferometric radii~\cite{Bozek:2013df}.
As will turn out, our results agree at a quantitative or semi-quantitative level with the 
experimental data for the highest centrality classes, supporting the collective picture 
of the most  central p-Pb collisions.
Our results also set the background for more elementary explanations of the correlation studies, 
based on saturation and the color-glass-condensate (CGC) theory~\cite{Dusling:2012iga,Dusling:2012wy,Dusling:2012cg,Kovchegov:2012nd,Kovner:2012jm,Dusling:2013oia}.
We note that certain features can also be obtained with cascade models~\cite{Lin:2004en,*Solanki:2012ne,*Xu:2011fe,*Burau:2004ev}.

We note that a certain degree of collectivity has  
been suggested for the p-p collisions as well~\cite{Luzum:2009sb,*d'Enterria:2010hd,*Bozek:2009dt,*CasalderreySolana:2009uk,*Avsar:2010rf,Bozek:2010pb,*Werner:2010ss}, 
where a same-side ridge is observed in the 2D correlations functions~\cite{Khachatryan:2010gv} for the highest multiplicity events.
This may indicate the presence of azimuthal correlations in the gluon emission from the initial 
state~\cite{Dusling:2012iga,Dusling:2012wy,Kovchegov:2012nd,Kovner:2012jm,Dusling:2013oia}. However, the same-side 
ridge observed in the p-p collisions could also result from the collective expansion of the 
created medium~\cite{Bozek:2010pb,*Werner:2010ss}. The intriguing questions concerning the p-p collisions will 
not be explored in this work, devoted to the detailed analysis of the p-Pb case.

Finally, we stress that our method is applicable to soft physics, related to particles produced with 
transverse momenta lower than, say, 2~GeV.

\section{The three-stage approach \label{sec:3}}

Our event-by-event approach is based on, by now, a standard picture involving three stages: 
generation of the initial densities, hydrodynamic evolution, and hadronization.
Certainly, variants of the modeling of each stage are present in the literature. We use 
the Glauber approach as implemented in GLISSANDO~\cite{Broniowski:2007nz} to model the initial phase, the 3+1~D viscous 
hydrodynamics~\cite{Bozek:2011ua}, and the statistical hadronization as implemented in THERMINATOR~\cite{Kisiel:2005hn,*Chojnacki:2011hb}. 

\subsection{Initial conditions \label{sec:init}}
 
The initial condition is generated with GLISSANDO~\cite{Broniowski:2007nz}, implementing various variants 
of the Glauber model~\cite{Glauber:1959aa,*Czyz:1969jg,*Bialas:1976ed,*Bialas:2008zza,Broniowski:2007ft}.
The parameters of the calculations are similar as in~\cite{Bozek:2011if}, except that they are adjusted for 
the collisions energy of $\sqrt{s_{NN}}=5.02$~TeV.
Thus we take the nucleon-nucleon (NN) inelastic cross section $\sigma_{NN}=67.7$~mb, moreover, we use a realistic 
(Gaussian) wounding profile~\cite{Rybczynski:2011wv} for the NN collisions. 

In the Glauber model, when a NN collision occurs, a {\em source} is produced, meaning a deposition 
of energy in a location in the transverse plane and spatial rapidity. 
In the standard wounded nucleon model it is assumed that a point-like source is located at the center 
of each participating nucleon, which leads to a rather large initial transverse size in the p-Pb system. 
Locating the source in the center-of-mass of the NN pair is also a possible model choice; it leads to
smaller initial distributions. In the results presented below we use both variants of the model, 
termed {\em standard} and {\em compact}. That way we may estimate the uncertainty related to modeling the initial 
phase within the Glauber treatment.

There is another important effect in the initial stage that influences the results: The weight of each source fluctuates according to some
statistical distribution, simply reflecting the fact that each NN collision may produce a different number of partons 
and therefore lead to varying deposition of the entropy. In the simulations~\cite{Broniowski:2007nz}, this feature is achieved by overlaying a suitable distribution 
of strength $w$ over the spatial distribution of the participant nucleons. 
As described in Sect.~\ref{sec:NB}, the observed multiplicity distributions can be described as convolution of the number of participant nucleons and a 
negative-binomial distribution. At the stage of the formation of the initial fireball it is equivalent to imposing fluctuations of the entropy
deposited per participant nucleon 
following the $\Gamma$ distribution, 
\begin{equation}
P_\Gamma(w)=\frac{w^{\kappa-1}\kappa^\kappa}{\Gamma(\kappa)}e^{-\kappa w} \label{eq:Gamma}
\end{equation}
with $\kappa=0.9$, as it leads to correct multiplicity distribution of the produced particles, cf.~Sec.~\ref{sec:NB}.
This case is labeled {\em Glauber+NB} as it eventually gives the
 multiplicity distribution as a  convolution of the Glauber Monte-Carlo 
distribution of participant
 nucleons and a negative 
binomial distribution. It is the most physical one and leads to best results when compared to the experiment. 
By construction, in the Glauber+NB case we place the sources at the centers of the participant nucleons, as in the standard case.

In all cases, after generating the spatial positions of sources, we smooth them with a Gaussian profile of width 0.4~fm. 
This physical effect (the sources do have a non-zero width) is also essential for 
hydrodynamics, which requires sufficient smoothness of the initial conditions. 
The smoothed initial distribution is placed on the 3+1~D lattice with spacing of 0.15~fm and then 
the event-by-event hydrodynamics is run.

\begin{table}
\caption{The mean values and standard deviations of the basic characteristics of the 
initial distributions for the centrality class 0-3.4\%. \label{tab:dis}}
\begin{tabular}{|c|cc|cc|cc|}
\hline
 & \multicolumn{2}{c}{standard} & \multicolumn{2}{|c|}{compact} & \multicolumn{2}{|c|}{Glauber+NB} \\ \hline
 & mean & std. dev. & mean & std. dev. & mean & std. dev. \\ \hline
$\langle r^2 \rangle^{1/2}$ [fm] & 1.54 & 0.15 & 0.93 & 0.06 & 1.45 & 0.22 \\
$\epsilon_2$                     & 0.25 & 0.13 & 0.19 & 0.09 & 0.34 & 0.16 \\
$\epsilon_3$                     & 0.29 & 0.13 & 0.18 & 0.09 & 0.32 & 0.15 \\
 \hline
\end{tabular}
\end{table}

\begin{figure}
\includegraphics[angle=0,width=0.42 \textwidth]{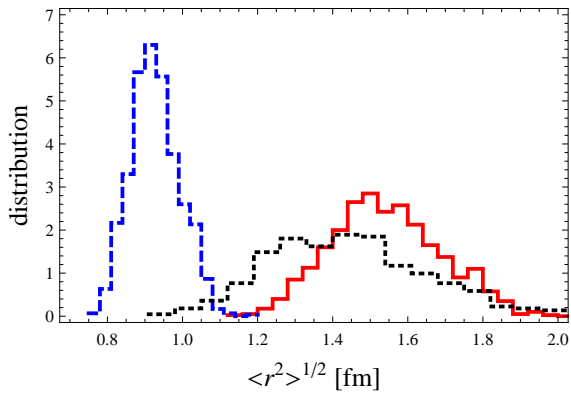} 
\caption{(Color online) The distribution of the transverse rms radius for the initial configuration 
in the standard case (solid line), compact case (dashed line), and the Glauber+NB case (dotted line), for the 
centrality class 0-3.4\%. 
\label{fig:sizedist}} 
\end{figure}
\begin{figure}
\includegraphics[angle=0,width=0.42 \textwidth]{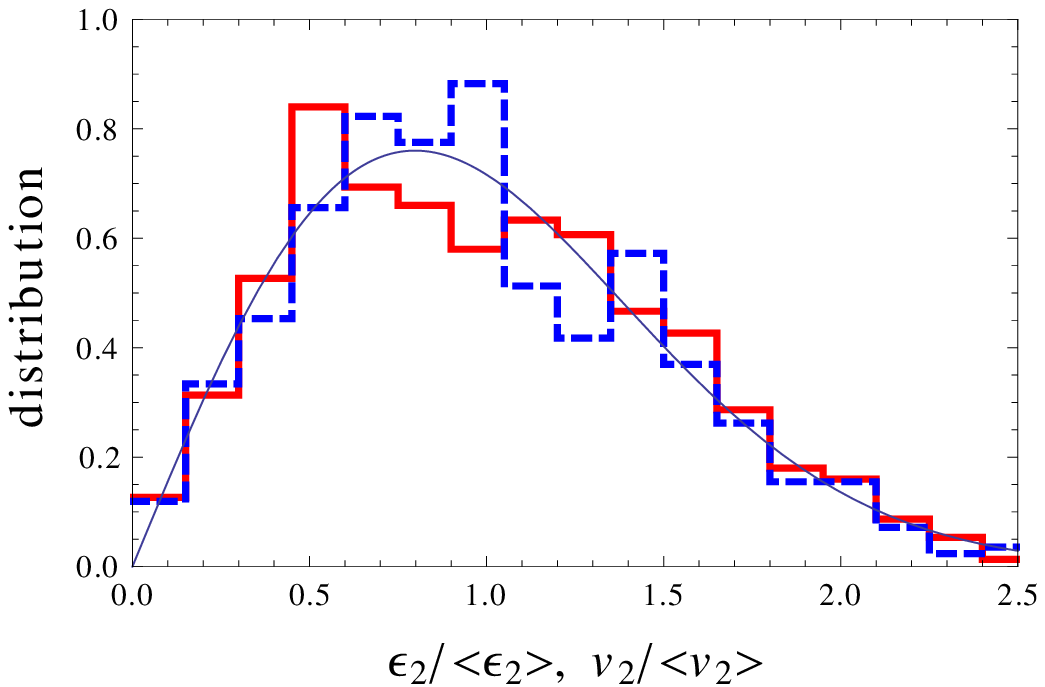} \\
\includegraphics[angle=0,width=0.42 \textwidth]{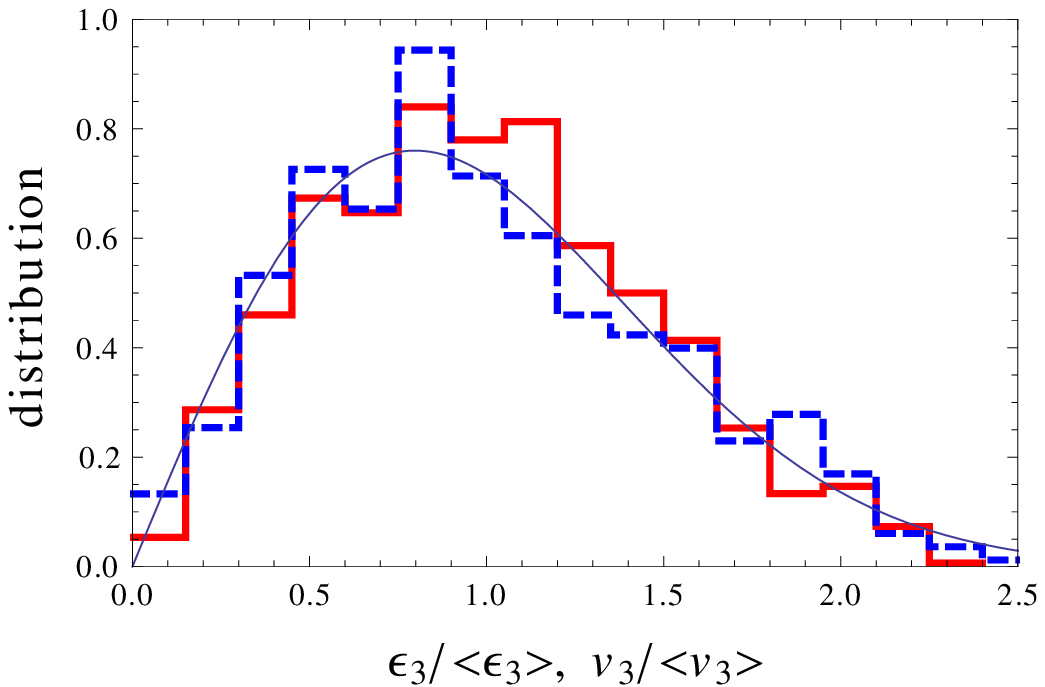} 
\caption{(Color online) The distributions of the scaled eccentricities $\epsilon_n$ and 
scaled harmonic flow coefficients $v_n\{2\}$, $n=2,3$ for the standard case. The thin solid line shows 
the Wigner distribution of Eq.~(\ref{eq:wig}). The flow coefficients $v_n\{2\}$ are discussed in the 
following sections.
\label{fig:v23dist}} 
\end{figure}

Some features of the resulting initial distributions are shown 
in Figs.~\ref{fig:sizedist} and \ref{fig:v23dist} for the collisions at the high centrality, $c$=0-3.4\%. 
We use here a few hundred configurations generated with GLISSANDO which are later fed
into the event-by-event hydrodynamic evolution. The basic properties of the distributions are listed in Table.~\ref{tab:dis}.

\subsection{Initial size \label{sec:size}}

At first glance, a rather surprising feature is the large transverse rms size of the initial distributions.
To understand, consider first the standard case, where the sources are located in the centers of the 
proton and of each of the participants from the lead nucleus.  
If the geometric ``hard-disk'' of radius $R$ were used for wounding, then the inelastic cross section 
would be $\sigma_{NN}=\pi R^2$, from where $R=1.47$~fm, corresponding to rms of 0.98~fm in a single NN collision.
This would be the uncertainty of the location of the point-like source created in the NN collision in the model.
However, we use the realistic Gaussian wounding profile~\cite{Rybczynski:2011wv} of the form
\begin{eqnarray}
p(b) = A \exp(-\pi A b^2/\sigma_{NN}), \;\; A=0.92. \label{gauprof}
\end{eqnarray}
After folding with the distributions of the nucleons in the Pb nucleus (via the Monte Carlo procedure in GLISSANDO) and after 
smoothing the positions with Gaussians of width 0.4~fm located in the centers of each source, we obtain the 
rms radius of the initial distribution listed in Table~\ref{tab:dis}, namely 1.54~fm.

In the compact case, where the the sources are placed in the center-of-mass of the colliding 
proton and the nucleon from the Pb nucleus, the source is significantly smaller, with the transverse rms of
0.93~fm. More involved models of the initial stage have been 
considered~\cite{Dumitru:2012yr,*Tribedy:2010ab,*Moreland:2012qw,*Schenke:2012hg}. The details of the energy deposition, such as the 
fluctuations and small scale structures, are relatively more important in  p-Pb than in A-A collisions. 
Our calculation, which uses two cases with significantly different 
initial size of the fireball can serve as an illustration of the effects of the
variation in initial size on the final harmonic flow observables.

\subsection{Initial eccentricities \label{sec:ecc}}

The participant eccentricities shown in Figs.~\ref{fig:v23dist} are defined in the usual way
for a given event as
\begin{equation}
\epsilon_n e^{i \Psi_n}= \frac{\int dx dy r^n \rho(x,y) e^{i\phi}}{\int dx dy r^n \rho(x,y)} \ ,
\end{equation}
where $\rho(x,y)$ is the initial transverse entropy distribution in the fireball at zero pseudorapidity and 
$\Psi_n$ is the azimuthal angle of the event plane. 
For the p-Pb system, the origin of the non-zero eccentricity lies in the fluctuations of the positions
of the participant nucleons. 
>From the formulas of Appendix~D of Ref.~\cite{Broniowski:2007ft} it is straightforward to obtain the result that for the very central collisions, $c=0$, 
where the average distribution in the Pb nucleus seen by the proton is azimuthally symmetric,
the scaled eccentricities $s=\epsilon_n/\langle \epsilon_n \rangle$ calculated from the positions of the
participant nucleons 
follow the Wigner distribution
\begin{eqnarray}
w(s)=\frac{\pi s}{2} \exp \langle - \frac{\pi s^2}{4} \rangle, \label{eq:wig}
\end{eqnarray}
independently of the rank $n$ of the harmonic component.
The distribution has $\langle s \rangle = 1$ and ${\rm var}(s)=4/\pi-1$~\cite{Broniowski:2007ft}. 
We note that this universality is clearly seen in Figs.~\ref{fig:v23dist}, up to the statistical noise 
and a slight departure from the central case of $c=0$. The eccentricities are calculated from the smooth
coarse-grained lattice distributions which introduces some small corrections with respect to the eccentricities calculated
from discrete positions of the the participant nucleons~\cite{Bhalerao:2011bp}.


\subsection{Pseudorapidity distribution \label{sec:rapidity}}

It is assumed that the initial transverse and longitudinal distributions are factorized. 
This assumption plays a key role in the interpretation of the development of the ridge structures 
in the hydrodynamic approach. It means that the transverse distribution is, within a reasonable 
range around the central region, independent of the pseudorapidity, i.e., approximately 
boost invariant. This leads to a correlation of ``geometry'' for the fireball slices separated by $\Delta \eta$, 
and, in consequence, to the correlation of flow. 
If an (approximately) boost-invariant fireball is formed, azimuthal correlations due to collective flow 
show up~\cite{Ollitrault:1992,Voloshin:2004th}. 

For the shape of the longitudinal distributions in the NN center of mass frame
we use the following profiles in the space-time rapidity $\eta_\parallel$:
\begin{eqnarray}
f(\eta_\parallel)_\pm &=& \exp\left(-\frac{(|\eta_\parallel|-\eta_0)^2}
{2\sigma_\eta^2}\theta(|\eta_\parallel|-\eta_0) \right) \times \nonumber \\
&& \frac{(y_b \pm \eta_\parallel) }{ y_b} \theta(y_b \pm \eta_\parallel), \label{eq:etaprofile}
\end{eqnarray}
with $\eta_0=2.5$, $\sigma_\eta=1.4$, and $y_b=8.58$ denoting the beam rapidity. The indices + and - correspond, respectively, to the 
distribution generated by the forward and backward moving participant nucleons. The same 
functional form of the profile has been successfully used 
in Refs.~\cite{Bozek:2010bi} to describe features of the A-A collisions, in particular the 
spectra in pseudorapidity and the behavior of the directed flow at RHIC. 
A phenomenological motivation for such ``triangular'' parameterizations has been 
discussed in~\cite{Bialas:2004su,*Nouicer:2004ke,*Gazdzicki:2005rr,*Bzdak:2009dr,Bozek:2010bi}.
The  resulting long-range correlations in pseudorapidity are strong in the asymmetric p-Pb collisions, hence the reaction planes
at different pseudorapidities are aligned (Sec. \ref{sec:scalar}).
To summarize, the initial conditions for hydrodynamics are the product of the smoothed transverse Glauber distribution in the transverse plane and 
the function~(\ref{eq:etaprofile}) in the longitudinal direction.

\subsection{Multiplicity distribution and fireball fluctuations \label{sec:NB}}

The Glauber Monte Carlo approach provides event-by-event fluctuations in the number of NN collisions and their distribution 
in the transverse plane. This mechanism explains most of the observed fluctuations in the shape of the fireball in the A-A
reactions~\cite{Alver:2007rm,Andrade:2006yh}. The event-by-event hydrodynamic expansion of the fluctuating fireball
generates azimuthally asymmetric flow and its fluctuations 
\cite{Andrade:2006yh,*Petersen:2010cw,*Alver:2010gr,*Qiu:2011hf,*Holopainen:2010gz,*Niemi:2012aj,Schenke:2010rr,Schenke:2012hg}. 
Some observables indicate that additional sources of fluctuations are present, beyond the fluctuations  in the 
number of participant nucleons, e.g., the event-by-event distribution of harmonic flow coefficients~\cite{collaboration:2011hfa,Gale:2012rq} or the 
multiplicity distributions~\cite{Dumitru:2011wq,Schenke:2012hg,Dumitru:2012yr}.

\begin{figure}
\begin{center}
\includegraphics[angle=0,width=0.42 \textwidth]{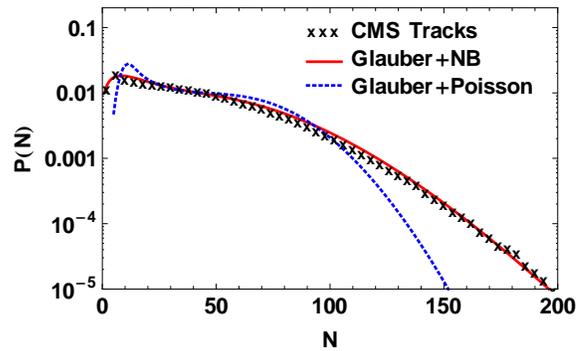} 
\end{center}
\vspace{-4mm}
\caption{(Color online) Multiplicity distribution of tracks with $p_\perp>0.4$~GeV and $|\eta|<2.4$ measured by CMS~\cite{cmswiki}. 
The dotted and solid lines denote the convolution of the distribution of participant nucleons from GLISSANDO with the Poisson and
negative binomial distributions, respectively. 
\label{fig:mult}} 
\end{figure}  

It is well known that the multiplicity distributions in the p-p collisions can be described by the negative binomial distribution~\cite{Giovannini:1985mz}
\begin{equation}
N_{\lambda,\kappa}(n)=\frac{\Gamma(n+\kappa)\lambda^n\kappa^\kappa}{\Gamma(\kappa)n! (\lambda+\kappa)^{n+\kappa}}, \label{eq:NB}
\end{equation}
where the multiplicity $n$ has the mean and variance given by  $\lambda$ and $\lambda(1+\lambda/\kappa)$, respectively.
Below we argue that in the p-Pb collisions we have a similar situation. 
In Sec.~\ref{sec:init} we stated that one should overlay a weight distribution over the spatial distribution of 
participant nucleons -- a feature implemented in GLISSANDO. Now we show how this distribution can be adjusted in such a way that the multiplicity data 
are properly reproduced. 

We use the multiplicity distributions of tracks observed in the
minimum-bias p-Pb collisions~\cite{cmswiki}.
In the three stage model of particle production, multiplicity fluctuations come from the
fluctuations of the initial entropy of the fireball, from the entropy production during the viscous hydrodynamic
stage, and from the statistical emission of hadrons at freeze-out. To a good approximation, in the considered regime of centralities the entropy after the hydrodynamic
expansion is directly proportional to the initial entropy, which reflects the fact that the deterministic hydrodynamic evolution does not 
introduce fluctuations~\cite{Olszewski:2013qwa}. 
For independent statistical emission, the number of emitted hadrons
is proportional to the final entropy and follows the Poisson distribution~\cite{Kisiel:2005hn,*Chojnacki:2011hb}.

First, we consider the case where there is no weight distribution overlaid over the participant nucleons.
Then the initial entropy is proportional to the number of participant nucleons and
the distribution of the observed tracks is a convolution of three distributions: the  distribution of participant
 nucleons, a Poisson distribution for each 
participant with a mean $\lambda$ defined as the average number of particles produced per participant, 
and a binomial distribution with success rate $p$ giving the probability of recording a track in the CMS acceptance. The folding yields
the multiplicity distribution of the produced hadrons of the form
\begin{equation}
P(n)=\sum_{i} P_{part}(i) \frac{\left( \lambda p i \right)^n e^{-\lambda p i}}{n!},
\end{equation}
where $P_{part}(i)$ is the distribution of the participant nucleons from the Glauber Monte Carlo.
The parameter $\lambda p=5.36$ is chosen to reproduce the mean number of the observed tracks. As we can see  
(the dotted line in Fig.~\ref{fig:mult}),  the multiplicity 
distribution from the Glauber model convoluted with the Poisson distribution 
is much too narrow and does not reproduce the experimentally observed high-multiplicity tail. 

The above shows that inserting a distribution of weights over the participant nucleons is necessary.
In that case the distribution of the observed tracks is a convolution of four distributions: the  distribution of participant nucleons, 
an overlaid distribution of weights, a Poisson distribution for the production of hadrons, 
and a binomial distribution for the experimental acceptance. When we use the distribution of weight in the form of the $\Gamma$ 
distribution~(\ref{eq:Gamma}), then its folding with the Poisson distribution yields the negative-binomial 
distribution~(\ref{eq:NB}), which we now take for hadrons produced per participant nucleon.
One finds 
\begin{equation}
P(n)=\sum_{i} P_{part}(i) N_{ p\lambda i,\kappa i}(n) .
\end{equation}
The experimental multiplicity distribution is now very well reproduced with the parameter values $\lambda p=5.36$ and $\kappa=0.9$
(solid line in Fig.~\ref{fig:mult}). We refer to this calculation as Glauber+NB.

The procedure outlined above is clearly a simplified picture of the multiplicity fluctuations in the 
relativistic nuclear collision. Further effects could be important, in particular
the shape of the multiplicity distribution can depend on 
the pseudorapidity window, the track acceptance of the CMS detector 
is not uniform, particles from jets that contribute to the the tails of the 
multiplicity distribution do not increase the fluctuations in the
thermalized fireball, or the entropy production in viscous hydrodynamics is not exactly linear.
Nevertheless, the considered mechanism of additional density fluctuations in the fireball can serve as a model to 
illustrate its expected effects on the eccentricities and the harmonic flow coefficients. Such effects in A-A collisions have been considered 
previously in the Glauber scheme~\cite{Broniowski:2007ft,Rybczynski:2008zg} and found to be relevant.


\begin{figure}
\begin{center}
\includegraphics[angle=0,width=0.31 \textwidth]{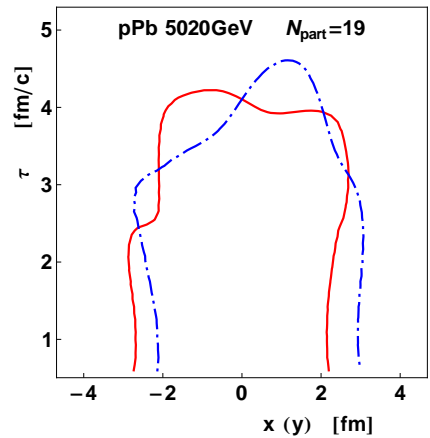} 
\end{center}
\vspace{-7mm}
\caption{(Color online) A sample event evolution, visualized via the freeze-out 
isotherms in the $x-\tau$ plane (solid) and the $y-\tau$ plane (dashed). 
The standard case.
\label{fig:his}} 
\end{figure}   

\begin{figure}
\begin{center}
\includegraphics[angle=0,width=0.31 \textwidth]{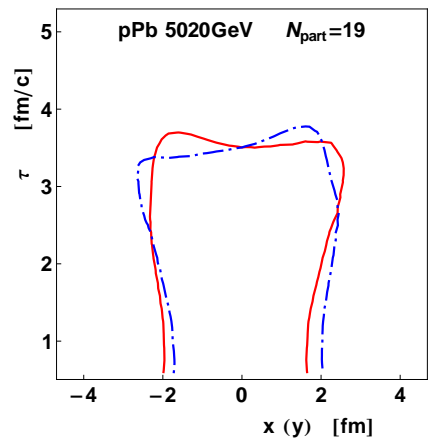} 
\end{center}
\vspace{-7mm}
\caption{(Color online) Same as Fig.~\ref{fig:his} for the compact-source case.
\label{fig:hisc}} 
\end{figure}   

\subsection{Hydrodynamics \label{sec:hydro}}

The hydrodynamic model used in this work is described in detail in~\cite{Bozek:2011if}. It carries out 
an event-by-event 3+1~D evolution and includes the shear and bulk viscosities.
The multiplicity expected in central p-Pb collisions is extrapolated linearly in the number of the participating nucleons from
the minimum bias results of the ALICE collaboration~\cite{ALICE:2012xs}.
That way the average initial entropy per participant 
is adjusted.
The shape of the entropy distribution follows the distribution of sources described in the previous section.
The starting time of hydrodynamics is chosen to be $\tau_0=0.6$~fm/c for the standard case, but the evolution with the
choice $\tau_0=0.2$~fm/c is also studied.  The relatively short total duration of the collective expansion phase makes the  results more 
sensitive to this very early stage and, possibly, to some nonequilibrium transient behavior \cite{Heller:2011ju,*Berges:2013eia,*Florkowski:2013sk}. 
The ratio of the shear viscosity $\eta$ to entropy density $s$
is $\eta/s=0.08$  or $\eta/s=0.16$, while the ratio of the bulk viscosity $\zeta$ to $s$ in the hadronic phase 
is $\zeta/s=0.04$ \cite{Bozek:2009dw}. For each case (standard, compact, higher viscosity, lower initial time, or Glauber+NB)
we produce  initial configurations that are evolved event-by-event with hydrodynamics to obtain freeze-out hypersurfaces of 
the constant temperature $T_f=150$~MeV. 

Two typical evolution histories for the standard and compact case, both for $N_{\rm part}=19$, are depicted in Figs.~\ref{fig:his}-\ref{fig:hisc}, where we show 
isotherms at $T_f=150$~MeV in the $x-\tau$ plane (solid lines) and in the $y-\tau$ plane (dashed lines). We note that although 
the systems originally have different sizes, the spatial spread of the isotherms at later times is similar, about 5~fm. The evolution of the standard 
source is about 15\% longer than from the compact-source case. The radial flow is larger for the compact case, as the system is more squeezed initially. 
This leads to 20\% higher values of the average transverse momentum.

\begin{figure}
\begin{center}
\includegraphics[angle=0,width=0.49 \textwidth]{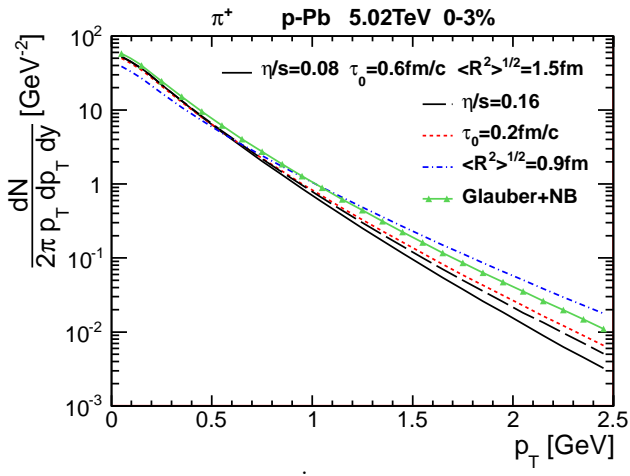} 
\end{center}
\vspace{-7mm}.
\caption{(Color online) The model predictions for the mid-rapidity transverse momentum spectra of $\pi^{+}$ for the most central 
events. The solid line is for the standard calculation, with the initial source rms size of $1.54$~fm, $\eta/s=0.08$, and
the initial time $\tau_0=0.6$~fm/c. The dotted line shows the results for the initial time of $0.2$~fm/c (and the other parameters unchanged), 
the dashed line stands for the calculations with $\eta/s=0.16$, and the dash-dotted line represents the calculation with the initial rms size $0.93$~fm/c. 
The solid line with the triangle symbols shows the Glauber+NB results, the case where  the Glauber Monte Carlo initial conditions are convoluted 
with the $\Gamma$ distribution (cf. Sec.~\ref{sec:NB}).
\label{fig:pionpt}} 
\end{figure}   

The transverse momentum spectra depend on the choice of initial conditions. The $p_T$-spectra for $\pi^+$ are shown in Fig.~\ref{fig:pionpt}. 
We notice that the spectra get harder with the increase of the 
shear viscosity, the decrease of the initial time $\tau_0$, or in the Glauber+NB case. The hardening of the spectra is most pronounced 
when starting the calculation from the compact source, which involves larger gradients present in the system.

\subsection{Statistical hadronization \label{sec:stat}}

For each freeze-out configuration we generate 1000 
THERMINATOR events to efficiently improve the statistics. 
This is a technical point, as physically one event should 
hadronize into one set of hadron distribution. The trick 
of running multiple THERMINATOR simulations on the same 
hydro event allows us to efficiently improve the statistics, as 
the computing  time for the 
hydro evolution is two orders of magnitude longer than 
for the generation of a single THERMINATOR event.

We note that the statistical hadronization built in THERMINATOR contains the non-flow correlations from all resonance decays. 
The use of a full Monte Carlo generator of hadron distributions is also of 
practical merits, as it allows implementation of the kinematic cuts, 
acceptance or efficiency from the experimental setup, 
which is crucial in comparisons to the data.

\subsection{Local charge conservation}

Sizable correlations among opposite-charge particles result from the local charge conservation~\cite{Bass:2000az}. 
There are indications that this effect is generated at hadronization~\cite{Jeon:2001ue,*Bozek:2003qi,*Aggarwal:2010ya,Bozek:2012en}, i.e., at the
late stage of the reaction. Our implementation of the charge balancing is based on the assumption that 
the particle-antiparticle  pairs of charged hadrons are emitted locally at 
freeze-out, carrying thermal distribution. The mechanism is described in detail in Ref.~\cite{Bozek:2012en}. 

\subsection{Transverse-momentum conservation \label{sec:cons}}

In our studies of the correlation variables we enforce 
the global transverse momentum conservation, which is important 
in correlation analyses~\cite{Borghini:2000cm,*Bzdak:2010fd}. 
In particular, it affects the shape of the two-particle correlations 
in relative pseudorapidity and azimuth. 
To satisfy the constraint approximately we require
the following condition on the global transverse momentum:
\begin{equation}
\sqrt{\left( \sum_{i} p_x \right)^2 + \left( \sum_i p_y \right)^2}< P_T, 
\end{equation}
where $i$ labels particles in the event.
We have found numerically that in the central p-Pb system it suffices to take $P_T = 5-10$~GeV.
That way we retain 5-10\% of the least-$P_T$ events from our full sample. A further 
lowering of $P_T$ does not affect the correlation results, while it deteriorates the statistics.
The momentum conservation is imposed when calculating the di-hadron correlation functions. For the calculation 
of the elliptic and triangular flow, imposing the momentum conservation in that form is irrelevant.

\subsection{Centrality definition}

The simplest determination of the centrality classes in our model can be obtained from conditions on the number of 
the participant nucleons. The collisions with $N_{part}\ge 18$ amount to $3.4\%$ of the most central events
from GLISSANDO. The next most central class is defined as $16 \le N_{part} \le 17$, which forms
centrality $3.4-7.8\%$. On the other hand, as noted in \cite{Bozek:2011if}, simplistic centrality definitions based on 
the impact parameter are ill defined for  central p-Pb collisions.

For the Glauber+NB case we define the centrality classes not by the number of participant nucleons, but through 
the total initial entropy in the fireball, i.e., we take into account the fluctuations from the overlaid $\Gamma$ distribution.
A more accurate determination, following closely to the experimental setup, should impose the cuts on the final multiplicity of the 
produced particles, instead of on $N_{part}$ or the initial entropy. 

\section{Two-particle correlations \label{sec:2D}}

\subsection{Definitions \label{sec:def}}

The basic objects of the study of this section 
are the two-dimensional two-particle correlation functions in relative pseudorapidity and azimuth.
These quantities in comparison to the CMS data~\cite{CMS:2012qk} have already been analyzed in~\cite{Bozek:2012gr}. 
Here we extend this analysis, comparing to the ATLAS data~\cite{ATLAS:pPb} as well.  

The simplest definition of the correlation function in the considered kinematic variables is
\begin{eqnarray}
\label{eq:2pc_s} 
C(\Delta\eta,\Delta\phi) \equiv
\frac{\langle \frac{d^{2} N^{\rm{pair}}}{d\Delta\eta\, d\Delta\phi} \rangle_{\rm events}}
     {\langle \frac{d^{2} N^{\rm{pair}}}{d\Delta\eta\, d\Delta\phi} \rangle_{\rm mixed \ events}}.
\end{eqnarray}
If the correlations were absent, $C_(\Delta\eta,\Delta\phi)=1$, thus unity is a natural scale for this measure. 
This correlation is used by the ATLAS collaboration~\cite{ATLAS:pPb}.

The ``per trigger'' correlation function, used by the CMS collaboration, is defined as~\cite{CMS:2012qk}
\begin{eqnarray}
\label{eq:2pc} \hspace{-4mm}
&&C_{\rm trig}(\Delta\eta,\Delta\phi) \equiv \frac{1}{N}\frac{d^{2} N^{\rm{pair}}}{d\Delta\eta\, d\Delta\phi}
= B(0,0) \frac{S(\Delta\eta,\Delta\phi)}{B(\Delta\eta,\Delta\phi)}, \nonumber \\
\end{eqnarray}
with $\Delta\eta$ and $\Delta\phi$ denoting the relative pseudorapidity and azimuth
of the particles in the pair. The signal and the mixed-event background are defined with the pairs from the same event,
and the pairs from the mixed events, respectively:
\begin{eqnarray}
\label{eq:s}
&& S(\Delta\eta,\Delta\phi) = \langle \frac{1}{N}\frac{d^{2}N^{\rm{pair}}}{d\Delta\eta\, d\Delta\phi} \rangle_{\rm events} , \\
&& B(\Delta\eta,\Delta\phi) = \langle \frac{1}{N}\frac{d^{2}N^{\rm{pair}}}{d\Delta\eta\, d\Delta\phi} \rangle_{\rm mixed \ events}. \nonumber
\end{eqnarray}
The number of particles $N$ in the prefactor (denoted by CMS as $N_{\rm{trig}}$) is the number of charged particles in 
a given centrality class and acceptance bin, corrected for the detector efficiency. The 
multiplication with the central bin content $B(0,0)$
in Eq.~(\ref{eq:2pc}) brings in the interpretation of Eq.~(\ref{eq:2pc}) as the average number of correlated pairs per 
trigger particle. 

To make quantitative comparisons easier, one also uses the projected correlation functions. A function used by the ATLAS collaboration is 
defined as 
\begin{eqnarray}
 Y(\Delta \phi)=\frac{\int B(\Delta \phi) d(\Delta \phi)}{\pi N} C(\Delta\phi) - b_{\rm ZYAM}, \label{eq:Y}
\end{eqnarray}
where $S(\Delta \phi)$ and $B(\Delta \phi)$ are averages of $S(\Delta \eta, \Delta \phi)$ and $B(\Delta \eta, \Delta \phi)$ over the 
chosen range in $\Delta \eta$ avoiding the central region, in particular $2< |\Delta \eta| < 5$ in the ATLAS analysis, and the constant 
$b_{\rm ZYAM}$ is such that the minimum of $Y(\Delta \phi)$ is at zero.

\subsection {Comparison to the ATLAS data}

\begin{figure}
\begin{center}
\includegraphics[angle=0,width=0.5 \textwidth]{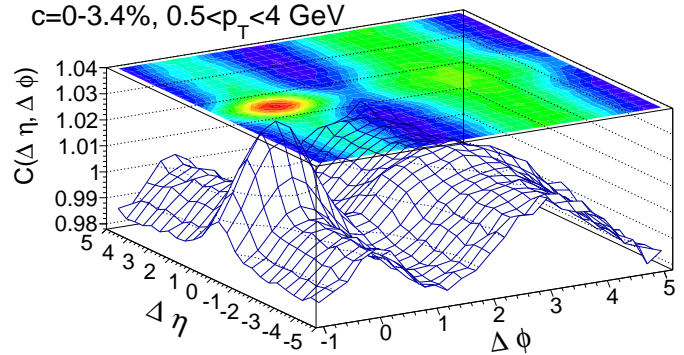} 
\end{center}
\vspace{-7mm}
\caption{(Color online) The charge-independent correlation function $C(\Delta \eta, \Delta\phi)$. The total transverse momentum is approximately 
conserved with the condition $P_T<10$~GeV. Charge balancing is included.
\label{fig:atlas}} 
\end{figure}   

The result of our simulations for the most central p-Pb collisions ($c=0-3.4$~\%) with the kinematic cuts 
corresponding to the ATLAS setup~\cite{ATLAS:pPb} is shown in Fig.~\ref{fig:atlas}. We display the standard-source case, as for the compact or 
Glauber+NB cases the 
results are quantitatively similar. We note the two prominent ridges, generated with flow, as well as the central peak, coming in our simulation from the charge 
balancing~\cite{Bozek:2012en}. 

The same-side ridge appears naturally as a consequence of the collective flow. More precisely, in our framework the shape and flow in the fireball 
in the forward and backward rapidity regions is correlated, reflecting the assumption on the factorization of the transverse and longitudinal distributions in the initial 
condition. In particular, the principal axes of the elliptic flow are correlated along the whole pseudorapidity span. Thus, there are more pairs with $\Delta \phi \sim 0$ and 
$\Delta \phi ~\sim \pi$ regardless of $\Delta \eta$. This ``flow explanation''  of the ridge formation is appealing in its simplicity.  

\begin{figure}
\begin{center}
\includegraphics[angle=0,width=0.48 \textwidth]{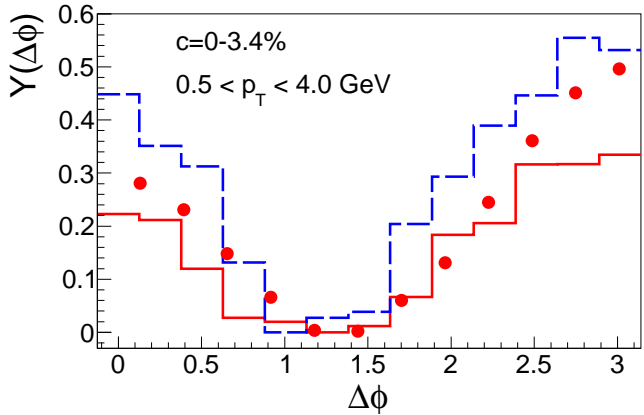} 
\end{center}
\vspace{-7mm}
\caption{(Color online) Projected and ZYAM-subtracted correlation function $Y\Delta(\phi)$ for the most central p-Pb collisions
for the standard source (solid line) and compact source (dashed line), compared to the ATLAS data (points) at 
$\Sigma E_T^{Pb} > 80$~GeV. The total transverse momentum is approximately 
conserved with the condition $P_T<5$~GeV. Charge balancing is imposed.
\label{fig:atzyam}} 
\end{figure}   

Next, to compare quantitatively to the data, we look at the projected correlation function $Y(\Delta \phi)$. There is a technical issue 
which must be discussed. By construction, the prefactor of $Y(\Delta\phi)$ is proportional to $\langle N(N-1) \rangle / \langle N \rangle$ -- the 
ratio of the average number of pairs to the average number of particles. Thus to reproduce $Y(\Delta\phi)$ in a model calculation  
one needs to have proper correlations, but also correct fluctuations of the multiplicity. The second requirement is not easy to 
accomplish, in particular as ATLAS is using the transverse energy to define the centrality classes, and the fluctuations of multiplicity are large. 
For that reason in our comparison we rescale our model $Y(\Delta\phi)$ in such a way that the subtraction constant $b_{\rm ZYAM}$ is the same in the model and in 
the experiment. This assumes that the mechanism generating the flow and the ridge structures is ``factorisable'' from the multiplicity fluctuations. 

The result of this procedure is shown in Fig.~\ref{fig:atzyam} for the most central collisions. We note that experimental data fall within our model results for the 
standard (solid line) and compact (dashed line) sources. We note that the compact source, leading to larger flow, has more prominent ridges.

The CMS correlation data for the p-Pb collisions have been compared to in our previous short paper~\cite{Bozek:2013df}, hence we do not repeat 
these results here, but only mention they are in semiquantitative agreement with the data.

\subsection{Effects of transverse momentum conservation}

\begin{figure}
\begin{center}
\includegraphics[angle=0,width=0.48 \textwidth]{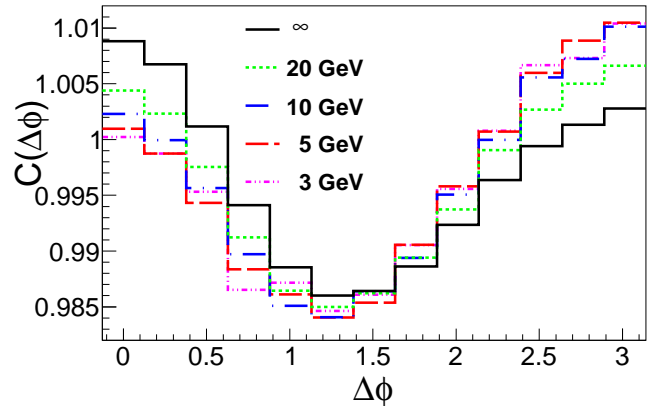} 
\end{center}
\vspace{-7mm}
\caption{(Color online) Projected correlation function $C\Delta(\phi)$ for the most central p-Pb collisions
for the standard source and several values of the maximum total transverse momentum $P_T$, listed in the legend.
\label{fig:PT}} 
\end{figure}

We can now demonstrate the relevance of the transverse momentum conservation and the simple procedure introduced in Sec.~\ref{sec:cons}. 
We use the projected correlation function $C(\Delta\phi)$ for that purpose. We note that limiting the value 
of the maximum total transverse momentum $P_T$ in the accepted events moves the strength from the same-side ridge to the 
away-side ridge. This is natural, as the momentum conservation increases the back-to-back motion of the particle. We note that 
for a practical purpose it is enough to use $P_T<5-10$~GeV. A further reduction changes the results very little at the expense of deteriorating 
the statistics. The numerical results, displaying the mentioned convergence, are shown in Fig.~\ref{fig:PT}.

\section{Harmonic flow \label{sec:flow}}

\subsection{Cumulant method}

>From the two- and four-particle cumulant method~\cite{Borghini:2000sa,*Bilandzic:2010jr} we obtain the values of the flow coefficients 
collected in Table~\ref{v:tab}. The kinematic cuts correspond to the ATLAS experimental setup. We compare the standard and the Glauber+NB 
simulation, without or with the local charge conservation, and list the results for the two highest centrality classes.
In Table~\ref{v2dep:tab} we show the dependence of $v_2\{2\}$ for the most central events on the 
parameters of the model: the value of the sheer viscosity $\eta$ and the time when hydrodynamics is initiated, $\tau_0$. 

\begin{table}[tb]
\caption{Model predictions for the elliptic and triangular flow coefficients from the cumulant method
for the p-Pb collisions at $\sqrt{s_{NN}}=5.02$~TeV. 
The cuts $|\eta|<2.5$, $0.3 < p_T < 5$~GeV correspond to the ATLAS setup. 
The standard and Glauber+NB cases are used, with $\eta/s=0.08$ and $\tau_0=0.6$~fm/c, without and with 
charge balancing. The errors are statistical and reflect the accumulated number of simulated THERMINATOR events (the actual error are somewhat 
larger due to a small number of the event sample). \label{v:tab}} 
\begin{tabular}{|r|rr|}
\hline
        &  $c$=0-3.4\% &  $c$=3.4-7.8\%         \\   \hline
  \multicolumn{3}{|c|}{standard, no balancing}  \\ \hline
$v_2\{2\}^2$ [$10^{-3}$]  & 3.70(1)     &   3.78(2)        \\
$v_3\{2\}^2$ [$10^{-3}$]  & 1.04(1)     &   0.95(1)        \\
$v_2\{4\}^4$ [$10^{-6}$]  & -0.4(4)     &   1.83(5)        \\
$v_3\{4\}^4$ [$10^{-6}$]  & 0.0(2)    &   -0.3(3)        \\
\hline
  \multicolumn{3}{|c|}{Glauber+NB, no balancing}   \\ \hline
$v_2\{2\}^2$ [$10^{-3}$] & 8.18(12)   & 8.24(10)   \\
$v_3\{2\}^2$ [$10^{-3}$] & 1.52(8)    & 1.51(6)    \\
$v_2\{4\}^4$ [$10^{-6}$] & 15(7)      & 16(6)      \\
$v_3\{4\}^4$ [$10^{-6}$] & -2(2)      & -2(2)      \\
\hline
  \multicolumn{3}{|c|}{Glauber+NB, with balancing}  \\ \hline
$v_2\{2\}^2$ [$10^{-3}$] &  8.22(7)   &  8.68(6)    \\
$v_3\{2\}^2$ [$10^{-3}$] &  1.57(4)   &  1.62(4)    \\
$v_2\{4\}^4$ [$10^{-6}$] &  19(4)     &   19(4)     \\
$v_3\{4\}^4$ [$10^{-6}$] &  -1(1)     &   0(1)      \\
\hline
\end{tabular}
\end{table}

\begin{table}[tb]
\caption{Parameter dependence of the predictions for the elliptic and triangular 
flow coefficient from the two-particle cumulant method for the p-Pb collisions at $\sqrt{s_{NN}}=5.02$~TeV, $c=0-3.4$~\%, $|\eta|<2.5$, $0.3 < p_T < 5$~GeV. 
Charge balancing not included. 
\label{v2dep:tab}}
 
\begin{tabular}{|r|rr|}
\hline
                & $v_2\{2\}$~[\%]  & $v_3\{2\}$~[\%]    \\
\hline
    \multicolumn{3}{|c|}{standard}  \\ 
\hline
$\eta/s=0.08$, $\tau_0=0.6$~fm/c &   6.09(1)  & 3.22(2)   \\
$\eta/s=0.08$, $\tau_0=0.2$~fm/c &   7.44(1)  & 4.49(1)   \\
$\eta/s=0.16$, $\tau_0=0.6$~fm/c &   5.57(1)  & 2.67(2)   \\
$\eta/s=0.16$, $\tau_0=0.2$~fm/c &   7.12(2)  & 4.01(2)   \\ 
\hline
    \multicolumn{3}{|c|}{Glauber+NB}  \\ 
\hline
$\eta/s=0.08$, $\tau_0=0.6$~fm/c &   9.0(1)    & 3.9(2)    \\ \hline
\end{tabular}
\end{table}
 
We note several qualitative features from Tables~\ref{v:tab} and \ref{v2dep:tab}:

\begin{itemize}
 \item The dependence on centrality is very weak, as expected from the flow generated mainly by the fluctuations of the initial condition. 
 \item The $v_n\{4\}^4$ coefficients are, within the statistical limit of our simulations, 
       compatible with zero for the standard case, while for the Glauber+NB simulations $v_2\{4\}^4$ is positive. This again 
       shows the fluctuation nature of the generated flow from the Glauber initial conditions. Additional fluctuations of the entropy deposited initially
       in the fireball increase the eccentricity and yield a nonzero value of $v_2\{4\}$ (cf. Table~\ref{tab:dis}).
 \item Increased sheer viscosity quenches, as expected, the flow. The relative effect is stronger for higher harmonics.
 \item A shorter time of starting hydrodynamics increases the flow, which again is expected.
 \item The effect of the local charge balancing increases somewhat the flow coefficients.
\end{itemize}

As a matter of fact, the first two items above are crucial for the proper interpretation of the observed 
phenomenon. Detailed comparisons of the model predictions to experimental measurements provide a way of learning about 
the shape and fluctuations of the initial density in the p-Pb system. The observation of nonzero $v_2\{4\}$ by the
ATLAS Collaboration~\cite{Aad:2013fja} indicates that in the small interaction region formed in the p-Pb collisions 
the large fluctuations of the energy deposited in each NN collision, as present in the Glauber+NB case, are crucial.
Thus the initial conditions from the Glauber model in p-Pb collisions are fluctuation-dominated, analogously to the central A-A collisions. 
The same observation applies to the final elliptic and triangular flow in p-Pb collisions.


\begin{figure}
\begin{center}
\includegraphics[angle=0,width=0.41 \textwidth]{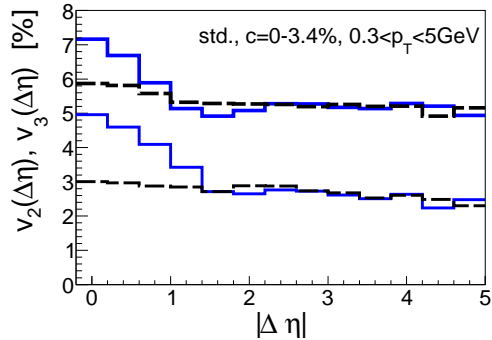} 
\end{center}
\vspace{-4mm}
\caption{(Color online) The flow coefficients $v_2(\Delta \eta)$ (upper lines) and $v_3(\Delta \eta)$ (lower lines) calculated
from the two-particle correlations as function of the relative pseudorapidity of the particles in the pair.
The solid and dashed lines are for the unlike- and like-sign pairs, respectively. The central peak is due to 
charge balancing and, to a lesser extent, resonance decays.
\label{fig:v23}} 
\end{figure}  

In view of the recent  experimental results  for the 2D correlations functions in p-Pb collisions, it is
interesting to look at the possibility of measuring directly the harmonic flow coefficients. We plot the elliptic
and triangular flow coefficients as functions of the pseudorapidity 
gap in Fig.~\ref{fig:v23}. The quantities are obtained in our hydrodynamic model 
from the Fourier decomposition of the correlation function $C_{\rm trig}(\Delta \eta, \Delta \phi)$~\cite{Bozek:2012en}. 
The non-flow effects present in our model are important only for pairs 
of small pseudorapidity separation. In the intervals $|\Delta \eta|>2$ the non-flow effects
from the resonance decays and the local charge conservation can be neglected. We note that the 
flow coefficients in Fig.~\ref{fig:v23} are sizable, thus
could be measured. It must be noted, however, that other sources of non-flow correlations  may be present
also in that kinematic region, but with smaller amplitudes, as measured in the p-p collisions~\cite{Khachatryan:2010gv}.

\begin{figure}
\begin{center}
\includegraphics[angle=0,width=0.5 \textwidth]{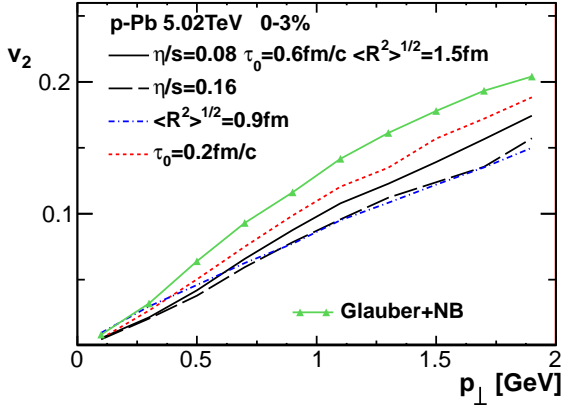} 
\end{center}
\vspace{-4mm}
\caption{(Color online) The $p_\perp$ dependence of the elliptic flow coefficient of charged particles with $|\eta|<1$, obtained from the second-order
cumulant method. The solid line corresponds to the standard calculation, $\eta/s=0.08$, $\tau_0=0.6$~fm/c, with the initial source rms size of $1.5$~fm,
the dotted line shows the case where the initial time is reduced to $0.2$~fm/c, the dashed line stands for the 
calculations with increased viscosity, $\eta/s=0.16$, and the dash-dotted line represents the case of the compact source with rms size of $0.9$~fm/c. 
Finally, the solid line with the triangle symbols shows the Glauber+NB case (Sect. \ref{sec:NB}).
\label{fig:v2pt}} 
\end{figure}

\begin{figure}
\begin{center}
\includegraphics[angle=0,width=0.5 \textwidth]{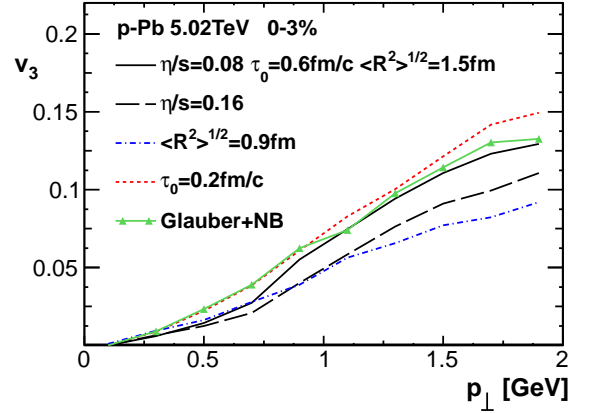} 
\end{center}
\vspace{-4mm}
\caption{(Color online) Same as Fig.~\ref{fig:v2pt} for the triangular flow coefficient, $v_3\{2\}$.
\label{fig:v3pt}} 
\end{figure}

The $p_\perp$-dependent  elliptic  and triangular flow coefficients calculated with the two-particle cumulant method 
\cite{Borghini:2000sa} are presented in Figs.~\ref{fig:v2pt} and \ref{fig:v3pt}. 
In the $p_\perp<2$~GeV range, where hydrodynamics 
applies, the flow coefficients show a typical hydrodynamic behavior and the magnitude of the flow is large. 
We find the elliptic (triangular) flow of about  10\% (5\%) for $p_\perp\sim 1$~GeV. 
The results are sensitive to the physical parameters of the model (cf. Table~\ref{v2dep:tab}). 
The flow decreases for larger viscosity or when using compact initial conditions, and increases when starting the hydrodynamic 
evolution earlier. It also increases with the presence of additional initial fluctuations, as in the Glauber+NB case. 
We notice a larger relative variation for the triangular flow than for the elliptic flow when varying the parameters.

\begin{figure}
\begin{center}
\includegraphics[angle=0,width=0.5 \textwidth]{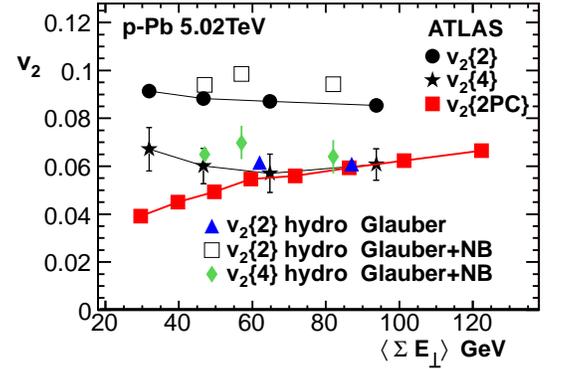} 
\end{center}
\vspace{-17mm}
\caption{(Color online) The elliptic flow coefficient of charged particles for $|\eta|<2.5$, $0.3<p_\perp<5.0$~GeV from the cumulant method 
$v_2\{2\}$ and $v_2\{4\}$, and from the di-hadron correlation function measured by the ATLAS Collaboration~\cite{Aad:2013fja}, compared to our hydrodynamic calculation for the standard case 
($v_2\{2\}$ at centralities 0-3.4\% and 3.4-7.8\%) and for the Glauber+NB case ($v_2\{2\}$ and $v_2\{4\}$ at centralities 0-5\%, 5-10\% and 10-20\%). 
The corresponding transverse-energy for the centralities in the model calculations is
obtained via interpolation of the experimental values.
\label{fig:v2atlas}} 
\end{figure}

\subsection{Scalar product method \label{sec:scalar}}

The correlation between particles produced in p-Pb collisions can have different origin. A way to reduce some of the 
non-flow contributions to the harmonic flow coefficients is to use methods involving a rapidity gap between the reference particles 
defining the event plane orientation and the particles used to calculate the flow coefficient. In this subsection we present results for the scalar
product method~\cite{Adler:2002pu,Luzum:2012da}. One defines the $Q_n$ vector 
\begin{equation}
Q_n^{A,B} e^{i\Psi_n}=\sum_k w_k e^{i n \phi_k}
\end{equation}
as a sum over charged particles in a given reference bin ($A$ or $B$). 
We use two definitions of the event plane, one with charged particles with $0.3<p_\perp<3$~GeV and $2.0<\eta<2.5$ (Pb side), 
or with $-2.5<\eta<2.0$ (proton side).  The weights are equal to the transverse energy ($w_k=E_\perp$) for
the $3.2<|\eta|<4.8$ bin. The resolution correction is
\begin{equation}
\bar{Q}_n^{A}=\sqrt{\frac{\langle Q_n^A Q_n^B \rangle \langle Q_n^A Q_n^C \rangle }{\langle Q_n^C Q_n^B \rangle}},
\end{equation}
where the reference bin $C$ is defined in all cases as $0.3<p_\perp<3$~GeV and $|\eta|<0.5$. We have checked that the results do
not differ noticeably when changing the $p_\perp$ or $\eta$ limits defining the $Q$ vectors.
The flow coefficients are then calculated as 
\begin{equation}
v_n^{A,B}\{SP\}=\frac{< Q_N^{A,B} \cos\left(n(\phi_k-\Psi_n)\right)>}{\bar{Q}_n^{A,B}} .
\end{equation}
The flow coefficients with reduced statistical error can be obtained with combined event planes on the proton and Pb sides.

\begin{figure}
\begin{center}
\includegraphics[angle=0,width=0.5 \textwidth]{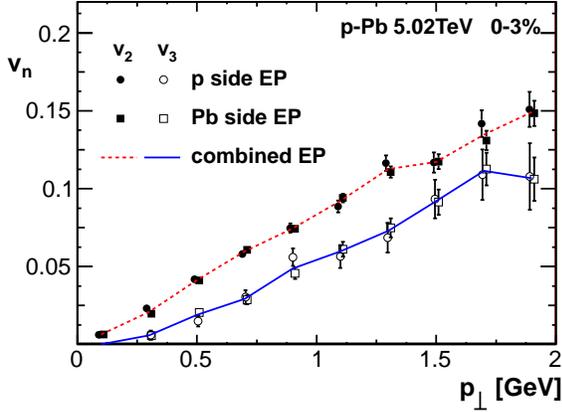} 
\end{center}
\vspace{-4mm}
\caption{(Color online) The elliptic (solid symbols and dashed line) and triangular (open symbols and solid line) flow coefficients 
obtained with the scalar product method.
The circles and squares represent the calculation using the Q vector calculated on the proton and lead side, respectively. The lines show
the result of combining the the two event planes. 
\label{fig:v2meth}} 
\end{figure}  

\begin{figure}
\begin{center}
\includegraphics[angle=0,width=0.5 \textwidth]{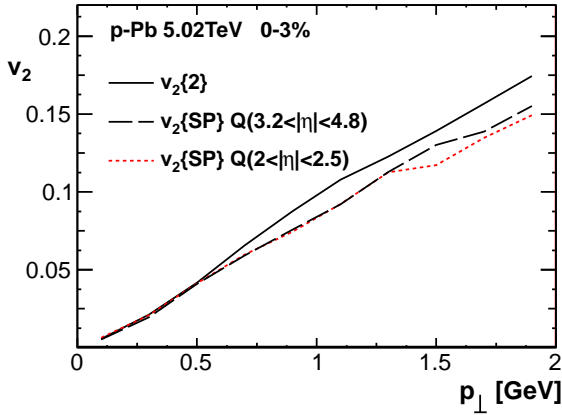} 
\end{center}
\vspace{-4mm}
\caption{(Color online) Comparison of the elliptic flow coefficients for $|\eta|<1$ calculated from the second cumulant method (solid line), from the scalar product method 
with event planes defined by charged particles in $2<|\eta|<2.5$ and $0.3<p_\perp<3.0$~GeV (dotted line), and from the scalar product method 
with event planes defined by the transverse energy in $3.2<|\eta|<4.8$  (dashed line).
\label{fig:v2comp}} 
\end{figure}  

\begin{figure}
\begin{center}
\includegraphics[angle=0,width=0.5 \textwidth]{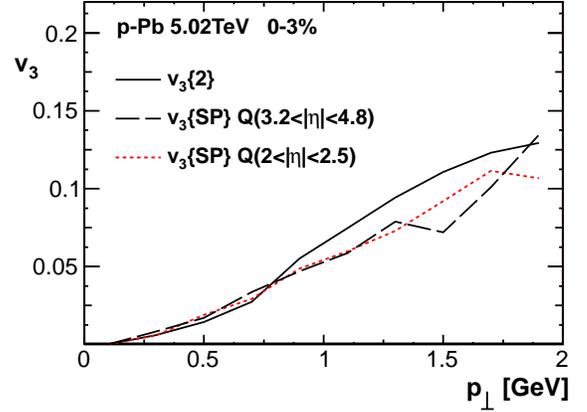} 
\end{center}
\vspace{-4mm}
\caption{(Color online) Same as Fig.~\ref{fig:v2comp} for the triangular flow.
\label{fig:v3comp}} 
\end{figure}

In Fig.~\ref{fig:v2meth} we show the elliptic and triangular flow coefficients obtained 
from the scalar-product method with the $Q$ vector from the bin $2.0<|\eta|<2.5$
on either the proton or the lead side. We notice that the two results are very consistent, with slightly smaller statistical errors for the 
$Q$ vector defined on the lead side. This reflects a better resolution of the event plane in that case. The azimuthally asymmetric 
initial source for hydrodynamic evolution is longitudinally extended, which yields a strong correlation between the event planes on 
the lead and proton sides. The observed two-particle correlation functions are almost symmetric for $\Delta \eta>0$ and $\Delta \eta<0$,
which shows that the correlations are similar on the proton and the lead side~\cite{CMS:2012qk}. The consistency of flow correlations 
defined with $Q$ vectors for positive and negative rapidities justifies the use of the combined $Q$ vector, which reduces the 
statistical error.

The results obtained with the $Q$ vectors defined by charged particle tracks or the calorimeter energy are compared to the results of the 
second cumulant method in Figs.~\ref{fig:v2comp} and \ref{fig:v3comp}. The elliptic and triangular flow coefficients obtained from the  
different definitions of the $Q$ vector are very similar. The second cumulant harmonic flow is calculated for smaller average 
pseudorapidity separation of the pair, thus contains some contribution of non-flow effects which increase the observed correlations. 
We expect that in the presence of additional non-flow correlation in the small system, such deviations could be larger.
By comparing the second cumulant $v_n$ to methods using large rapidity gaps, the importance of such non-flow correlations could 
be estimated in the data.

\subsection{Correlations of flow and the initial geometry \label{sec:ficor}}

\begin{figure}
\includegraphics[angle=0,width=0.48 \textwidth]{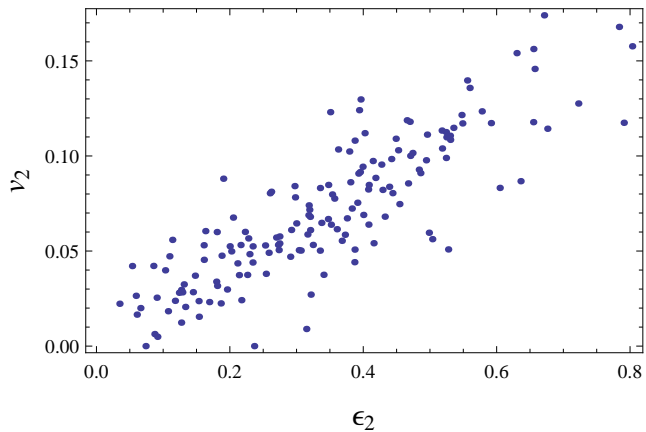} 
\caption{The scattered plot of the event-by-event eccentricity-elliptic flow correlations. Glauber+NB, correlation coefficient 0.85. \label{fig:e2v2}} 
\end{figure}

\begin{figure}
\includegraphics[angle=0,width=0.48 \textwidth]{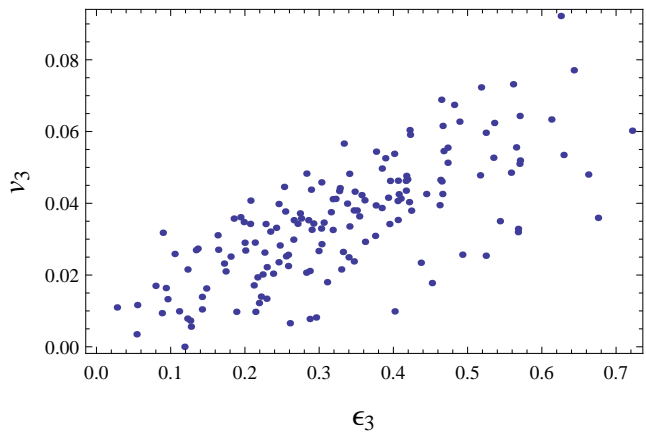} 
\caption{Same as Fig.~\ref{fig:e2v2} for the triangular case. Correlation coefficient 0.74.
\label{fig:e3v3}} 
\end{figure}

One of the main reasons to study the flow is to acquire the knowledge on 
the early phase of the reaction. One result that holds event-by-event is the proportionality of the 
eccentricity coefficients of the ``geometric'' distribution, $\epsilon_n$, to the coefficient 
of the harmonic flow of the produced hadrons, $v_n$. In Figs.~\ref{fig:e2v2} and \ref{fig:e3v3} we show 
the event-by-event scattered plots of eccentricity-flow distributions. For this calculation the 
hydrodynamic events are combined from $1000$ THERMINATOR events
corresponding to the same freeze-out configuration. We notice large correlation coefficients, 
defined as
\begin{eqnarray}
\rho=\frac{\langle \epsilon_n v_n\{2\} \rangle - \langle \epsilon_n \rangle \langle v_n\{2\}\rangle}{{\rm var}(\epsilon_n) {\rm var}(v_n\{2\})} 
\end{eqnarray}
in these distributions, $\rho=0.85$ for the elliptic and $\rho=0.74$ for the triangular case, respectively. This feature, well know for the A-A 
collisions, is vividly present in our treatment of the p-Pb collisions.

\section{Conclusion}

We have analyzed various aspects of soft collective dynamics of the relativistic p-Pb collisions in the approach consisting 
of three stages: Glauber modeling of the initial phase, event-by-event viscous 3+1~D hydrodynamic, and statistical hadronization. 
Our analysis shows that the collective dynamics may very well be present in the highest-centrality p-Pb system
formed in ultra-relativistic heavy-ion collisions. The application of the three-stage model, 
where the shape fluctuations in the initial stage are carried over to the harmonic flow coefficients 
in the hadronic spectra, allows for a quantitative understanding of the data for $v_2$ and $v_3$, as well 
as to describe the ridge structures in the two-particle correlation functions. The issues 
connected to the femtoscopic variables in p-Pb collisions, which display considerable sensitivity 
to collectivity, have been presented elsewhere~\cite{Bozek:2013df}.
 
Thus, following the successful experience of describing the A-A collisions in the three-stage approach, 
we argue that the collective scheme provides a natural and conventional explanation of numerous 
aspects of the soft dynamics of the ``small'' p-Pb system. 

In central p-Pb collisions, the initial shape eccentricity parameters $\epsilon_n$ are 
entirely due to fluctuations. These fluctuations are enhanced by the distribution of overlaid 
weights on the spatial distribution of the participant nucleons. We find it quite remarkable that the 
same distribution that explains the multiplicity distribution of the produced hadrons in minimum-bias 
collisions leads also to quantitative agreement for the values of the elliptic and triangular flow coefficients measured recently 
by the ATLAS collaboration~\cite{Aad:2013fja}. This agreement includes the elliptic flow coefficient obtained from the 
four-particle cumulants.

We argue that the lowest harmonic flow coefficient may be measured directly in the LHC p-Pb experiments, hence we 
compute them through various methods (cumulant, scalar-product, rapidity-gap).
These predictions, as well as the femtoscopic radii~\cite{Bozek:2013df},  will hopefully be verified shortly in the 
upcoming experimental analyzes.

\begin{acknowledgments}
Supported  by National Science Centre, grants DEC-2011/01/D/ST2/00772 and 
DEC-2012/05/B/ST2/02528, by PL-Grid Infrastructure, and by the European Research Council under the
Advanced Investigator Grant ERC-AD-267258.                                  
\end{acknowledgments}

\bibliography{hydr}

\end{document}